\documentclass[a4paper,12pt, epsfig]{article}
\usepackage{epsfig}
\usepackage{amssymb}
\usepackage{amsfonts}
\usepackage{amsmath}

\newskip\humongous \humongous=0pt plus 1000pt minus 1000pt

\newif\ifdtup

\catcode`\@=11

\@addtoreset{equation}{section}
\def\theequation{\thesection.\arabic{equation}}

\def\@normalsize{\@setsize\normalsize{15pt}\xiipt\@xiipt
\abovedisplayskip 14pt plus3pt minus3pt%
\belowdisplayskip \abovedisplayskip
\abovedisplayshortskip \z@ plus3pt%
\belowdisplayshortskip 7pt plus3.5pt minus0pt}

\def\small{\@setsize\small{13.6pt}\xipt\@xipt
\abovedisplayskip 13pt plus3pt minus3pt%
\belowdisplayskip \abovedisplayskip
\abovedisplayshortskip \z@ plus3pt%
\belowdisplayshortskip 7pt plus3.5pt minus0pt
\def\@listi{\parsep 4.5pt plus 2pt minus 1pt
     \itemsep \parsep
     \topsep 9pt plus 3pt minus 3pt}}

\relax

\catcode`@=12

\catcode`\@=11

\def\section{\@startsection{section}{1}{\z@}{3.5ex plus 1ex minus
   .2ex}{2.3ex plus .2ex}{\large\bf}}

\def\thesection{\arabic{section}}
\def\thesubsection{\arabic{section}.\arabic{subsection}}

\def\appendix{\setcounter{section}{0}
 \def\thesection{Appendix \Alph{section}}
 \def\thesubsection{\Alph{section}.\arabic{subsection}}
 \def\theequation{\Alph{section}.\arabic{equation}}}

\def\SymBoxes#1#2#3#4{\newdimen\un@t \un@t#3%
\raisebox{#1}{\rule{#2\un@t}{#4}\hskip-#2\un@t
\@tempdimb\un@t \advance\@tempdimb by-#4\@tempcntb#2\relax%
\@whilenum{\@tempcntb>0}\do{
\rule{#4}{\un@t}\hskip\@tempdimb \advance\@tempcntb by\m@ne}%
\hskip-#2\un@t \rule[\un@t]{#2\un@t}{#4}%
\rule[\un@t]{#4}{#4}\hskip-#4
\rule{#4}{\un@t}}\hskip-#4}                

\begin{document}

\newcommand{\beq}{\begin{equation}}
\newcommand{\eeq}{\end{equation}}
\newcommand{\bea}{\begin{eqnarray}}
\newcommand{\eea}{\end{eqnarray}}
\newcommand{\beas}{\begin{eqnarray*}}
\newcommand{\eeas}{\end{eqnarray*}}
\newcommand{\defi}{\stackrel{\rm def}{=}}
\newcommand{\non}{\nonumber}
\newcommand{\bquo}{\begin{quote}}
\newcommand{\enqu}{\end{quote}}
\renewcommand{\(}{\begin{equation}}
\renewcommand{\)}{\end{equation}}
\def\IZ{{\mathbb Z}}
\def\IR{{\mathbb R}}
\def\IC{{\mathbb C}}
\def\IQ{{\mathbb Q}}

\def\CM{{\mathcal{M}}}
\def\dCM{{\left \vert\mathcal{M}\right\vert}}

\def \d{\textrm{d}}
\def \p{\partial}

\def \pr{\prime}

\def\Tr{ \hbox{\rm Tr}}
\def\half{\frac{1}{2}}

\def \eqn#1#2{\begin{equation}#2\label{#1}\end{equation}}
\def\de{\partial}
\def\Tr{ \hbox{\rm Tr}}
\def\H{ \hbox{\rm H}}
\def\HE{ \hbox{$\rm H^{even}$}}
\def\HO{ \hbox{$\rm H^{odd}$}}
\def\K{ \hbox{\rm K}}
\def\Im{ \hbox{\rm Im}}
\def\Ker{ \hbox{\rm Ker}}
\def\const{\hbox {\rm const.}}
\def\o{\over}
\def\im{\hbox{\rm Im}}
\def\re{\hbox{\rm Re}}
\def\bra{\langle}\def\ket{\rangle}
\def\Arg{\hbox {\rm Arg}}
\def\Re{\hbox {\rm Re}}
\def\Im{\hbox {\rm Im}}
\def\exo{\hbox {\rm exp}}
\def\diag{\hbox{\rm diag}}
\def\longvert{{\rule[-2mm]{0.1mm}{7mm}}\,}
\def\a{\alpha}
\def\dag{{}^{\dagger}}
\def\tq{{\widetilde q}}
\def\p{{}^{\prime}}
\def\W{W}
\def\N{{\cal N}}
\def\hsp{,\hspace{.7cm}}
\newcommand{\C}{\ensuremath{\mathbb C}}
\newcommand{\Z}{\ensuremath{\mathbb Z}}
\newcommand{\R}{\ensuremath{\mathbb R}}
\newcommand{\rp}{\ensuremath{\mathbb {RP}}}
\newcommand{\cp}{\ensuremath{\mathbb {CP}}}
\newcommand{\vac}{\ensuremath{|0\rangle}}
\newcommand{\vact}{\ensuremath{|00\rangle}}
\newcommand{\oc}{\ensuremath{\overline{c}}}

\def\M{\mathcal{M}}
\def\F{\mathcal{F}}
\def\d{\textrm{d}}

\def\eps{\epsilon}

\begin{titlepage}


\begin{flushright}
SITP-09/05,
ITEP-TH-06/09,
TAUP-2895/09
\end{flushright}

\begin{center}
{\Large {\bf Chiral Symmetry Breaking with non-SUSY D7-branes in ISD backgrounds \\
}}
\end{center}

\vspace{1em}

\begin{center}
{\large  Anatoly Dymarsky$^1$, Stanislav Kuperstein$^2$ and Jacob Sonnenschein$^3$ }
\end{center}

\renewcommand{\thefootnote}{\arabic{footnote}}

\begin{center}

\vspace{1em}
{\em  { $^1$Stanford Institute for Theoretical Physics, \\ Stanford University,  Stanford, CA 94305, USA\\}}   \texttt{dymarsky@stanford.edu}

\vspace{0.5em}
{\em  { $^2$Theoretische Natuurkunde,
Vrije Universiteit Brussel \\ and The International Solvay Institutes,
Pleinlaan 2,  B-1050 Brussels, Belgium \\}} \texttt{skuperst@vub.ac.be}

\vspace{0.5em}
{\em  { $^3$School of Physics and Astronomy,
The Raymond and Beverly Sackler Faculty \\of Exact Sciences,
Tel Aviv University, Ramat Aviv, 69978, Israel \\}} \texttt{cobi@post.tau.ac.il}

\end{center}

\vspace{1.5em}

\noindent
\begin{center} {\bf ABSTRACT} \end{center}

Supersymmetric probe branes satisfy the $\kappa$-symmetry condition which ensures that the action is minimized
with respect to both variations of the embedding and of the world-volume gauge field. We observe that
the $\kappa$-symmetry condition for a $D7$-brane in an imaginary self-dual (ISD) background
can be generalized yielding minimization
of the action with respect to variations of the gauge field only but not of the embedding. This provides a new
way to construct non-BPS solutions for D7-branes once the embedding extremizes
the geometrical volume of the brane.
We then apply this method to the Klebanov-Strassler background and find a new $D7-\bar{D}7$ brane
configuration that realizes the spontaneous breaking of flavor chiral symmetry as evidenced by the Goldstone boson
identified in the spectrum. This result generalizes our previous
construction for the Klebanov-Witten model. We compare our setup with the Sakai-Sugimoto model and discuss
possible applications to QCD-like physics.


\end{titlepage}

\bigskip

\hfill{}
\bigskip

\tableofcontents

\setcounter{footnote}{0}
\setcounter{figure}{0}

\section{\bf Introduction}

The steady progress of the gauge/string duality \cite{Maldacena}
in describing four-dimensional phenomena gives rise to a hope to find a stringy (gravity) model for the QCD and hadron physics. Yet among other unresolved problems finding a suitable model to capture the spontaneous breaking of flavor chiral symmetry ($\chi$SB) proven to be a complicated task.
The fundamental matter appears as a result of incorporating the flavor branes \cite{Karch:2002sh} into the gravity background.
To find such a solution is a formidable task. The problem greatly simplifies
in the so-called quenched approximation \mbox{$N_{\rm c} \gg N_{\rm f}$} when one can neglect the back-reaction of flavor branes and consider them in probe approximation. Even in this case to find a solution could be complicated, especially if a non-trivial background NS-NS flux is present.

The study of the flavor chiral symmetry breaking took a new turn after Sakai and Sugimoto proposed a model
\cite{Sakai:2004cn} providing a successful holographic description of the phenomenon.
The Sakai-Sugimoto model is based on incorporating $D8$-branes into the background of non-extremal D4-branes. The former is also known as Witten's model \cite{Witten:1998zw}. The D8-branes have the U-like shape which can be interpreted as the stacks of D8 and anti D8-branes merging in the IR region. Since the  branes and the anti-branes are separated in the UV, the flavor symmetry  is $U(N_{\rm f})_L\times U(N_{\rm f})_R$. In the IR region though the flavor symmetry is broken to the diagonal $U(N_{\rm f})_D$ by a quark VEV. Besides the chiral symmetry breaking the Sakai-Sugimoto setup
provides a holographic realization of light mesons and baryons.

Despite the success in describing chiral symmetry breaking the Sakai-Sugimoto model suffers from the drawbacks inherited from the Witten's model. In particular it is inconsistent in the UV region because the string coupling diverges there.
Besides that the geometry includes a compact circle. Therefore the four dimensional description breaks down for the energies exceeding the compactification scale. A potential way to bypass these problems is to study the flavor branes in the  confining backgrounds with constant dilaton like the one found by Klebanov and Stassler \cite{KS}.
Motivated by the work of Sakai and Sugimoto we expect that to model the chiral symmetry breaking the flavor branes
have to be U-like shaped with each end preserving a distinct set of supercharges while the whole solution is not supersymmetric. One way to achieve this in the KS case is to wrap the D7-brane over the three-cycle $S^3$ at the base
of the conifiold $T^{1,1}\cong S^2\times S^3$ \cite{KW,EK} while D7-brane is also stretching in the rest of the confiold
$R^+\times S^2$ forming a U-like shape.

An important step in this direction was recently undertaken in \cite{CobiMe} where the proposed $D7$-$\bar{D}7$ setup was investigated in the case of the conformal Klebanov-Witten (KW) background \cite{KW}. It was found there that the asymptotic location of two branches are not antipodal on the $2$-sphere but have a fixed angle difference of $\sqrt{6}/4  \pi$. Because  the KW background is conformal the angle separation is independent of the only free parameter of the solution $r_0$, which marks the lowest point  of the D7-brane's profile. In the dual field theory $r_0$ corresponds
to a chiral symmetry breaking VEV. The corresponding Goldstone boson of the broken flavor chiral symmetry was also identified in the spectrum \cite{CobiMe}.

In an earlier attempt \cite{Sakai:2003wu} a similar configuration of $D7-\bar D7$ flavor branes was investigated in the confining background of the KS solution. Despite some interesting insights into the geometry of the D7-brane profile the solution was not completed because of complications due to the nontrivial background NS-NS flux of the Klebanov-Stassler
(KS) solution.
The main goal of this paper is to generalize the results of  \cite{CobiMe} to the Klebanov-Strassler background
thus completing and generalizing the construction of \cite{Sakai:2003wu}. As we have already mentioned above the main
obstacle in this case is the non-trivial background $B$-field which couples to the word-volume fields and excludes the configurations with vanishing gauge field.  Similarly to \cite{CobiMe} the configuration is question should break supersymmetry. This and the non-linear nature of the probe's equation of motion significantly complicate the process of finding the solution.

In the quest for a probe brane solutions in the KS model (or any other background) a remarkable
simplification can be achieved if one focuses only on the supersymmetric configurations.
In this case a non-linear second order equations of motion can be substituted by the first order $\kappa$-symmetry constraint \cite{kappasym} which ensures that the D7's action is minimized. Thus in the case of the ISD  background with constant dilaton the $\kappa$-symmetry requires the D7-brane to be embedded along a holomorphic cycle and the gauge invaraint world-volume gauge field \mbox{$\F \equiv \varphi^\star(B) + 2 \pi \alpha' F$} to be of
the $(1,1)$ type and anti-self dual (ASD):
\bea
\label{IntroASD}
\F = - \star_4 \F\ .
\eea
Although to satisfy these requirements may turn to be a complicated task \cite{ShOuyang} in general the supersymmetric solutions are easier to find. Unfortunately the known supersymmetric solutions \cite{Ouyang,Me,Levi:2005hh} have a
geometry of the single branch which looks topologically trivial in the UV and hence don't describe flavor chiral symmetry breaking in the dual gauge theory. Rather the flavor symmetry consists of a single $U(N_{\rm f})$.
The holomorphic embddings has been also extensively studied beyond the quenched approximation taking into account
the backreaction of the branes \cite{Un-quenched}.

Our primary observation, however, is that the ASD condition (\ref{IntroASD}) minimizes the D7-brane action
with respect to the world-volume gauge field even though the  other two criteria may not be met. Therefore it can be used to construct a non-supersymmetric  non-BPS solutions provided one can solve the equation for the embedding. In fact the task is even simpler since the ASD gauge field decouples from the equations governing the embedding leading to an intuitive problem of extremizing  geometrical volume of the D7-brane. In this paper we use this method to find a solution with the $SU(2)$ symmetry which  unambiguously fixes the embedding upon the boundary conditions in the UV are specified. The profile we found is not holomorphic, yet it minimizes the volume within its class of symmetries. We also complete the solution by constructing the ASD gauge field that satisfies the Bianchi identity.

The geometry of the solution is a reminiscent of the one found in the KW case  \cite{CobiMe}.
It is a U-shaped configuration which wraps $S^3$ at the base of the conifold and is stretched along the equator on $S^2$.
The only free parameter of the solution $\tau_0$ measures the distance between the lowest point of the profile and the tip of the conifold. For $\tau_0$ much larger than the deformation parameter of the conifold the geometry approaches the conformal one
and the angle separation between the D7-brane and the anti-brane approaches the KW value $\sqrt{6}/4 \pi$. For the configuration stretching all way to the tip $\tau_0=0$ the locations of brane and the anti-brane on $S^2$ are exactly antipodal.

The paper is organized as follows.
In  the next section we briefly review the $\kappa$-symmetry condition for the D7-brane and show that the ASD condition along minimizes the action with respect to the perturbations of the world-volume gauge field. We then proceed with formulating a method to find non-supersymmetric solutions via extremizing the geometrical volume of the embedding.
In Section \ref{D7KS} we construct such an embedding for the D7-brane in the KS background and accompany  it by an appropriate  ASD world-volume flux in Section \ref{ASDconditionKS}. In this way we construct the model with the sponteniously broken flavor chiral symmetry. We elaborate on the four dimensional physics of the model in Section \ref{dGT} and briefly discuss other possible applications of our findings in Section \ref{Discuss_s}. Various technical details are delegated to the appendices.

\noindent

\section{\bf General Idea}
\label{secGI}

Motivated by the Introduction we are interested in finding a solution for a D7-brane embedded in a type IIB background with a Poincare symmetry and a constant dilaton $e^{\phi}=g_s$. As discussed  in \cite{ISD-contDiltaton} such a background has
imaginary-self-dual 3-form flux $G_3 = F_3 + \frac{i}{g_s} H_3$ and is usually referred as an ISD background.
The ten dimensional geometry of an ISD background is a warped product
\begin{equation}
\label{metric10}
\d s_{(10)}^2 = h^{-1/2} \left( \d x_0^2 + \ldots + \d x_3^2 \right) + h^{1/2} \d s_{M_6}^2,
\end{equation}
of the flat Minkowski space and a $6d$ dimensional CY space $M_6$ with the metric $\d s_{M_6}^2$.
The warp factor $h$ depends only on the coordinates along $M_6$ (radius $\tau$ in the conifold case)
and determines the  self-dual RR $5$-form
\begin{equation}
\label{F5C4}
\widetilde{F}_5 = \left(1 + \star_{10} \right) \d C_4 \qquad \textrm{where} \qquad
C_4 = h^{-1} \cdot \d x_0 \wedge \ldots \wedge \d x_3.
\end{equation}
The complex $3$-form $G_3$ is zero in the Minkowski space and
is imaginary self dual $G_3 = \star_6 i G_3$ with respect to  the $6d$ metric on $M_6$. Providing that the $3$-form is of the $(2,1)$ type the background has four supercharges \cite{ISD-contDiltaton} (see \cite{ISD-NONcontDiltaton} for the non-constant dilaton case)
and can be dual to a certain $4d$  $\mathcal{N}=1$ gauge theory.

In addition the self-duality of $G_3$ and the explicit form of $C_4$ in (\ref{F5C4})
imply that $F_7$ defined as
\begin{equation}
\label{F7}
F_7 = \star_{10} F_3 + C_4 \wedge H_3   \qquad \textrm{with} \qquad
  \star_{10} F_3 = h^{-1} \star_6 F_3 \wedge \d x_0 \wedge \ldots \wedge \d x_3
\end{equation}
identically vanishes \cite{Me}.

A particularly well known example of such a background is the Klebanov-Strassler solution \cite{KS}. Although the remnant of this section is applicable to any ISD background one can have  KS background in mind in what follows.
We particularly focus on the KS background starting Section \ref{D7KS} where we construct a non-SUSY
D7-brane illustrating the main idea of this section.

Now let us focus on a D7-brane that spans the Minkowski space together with a \mbox{$4$-cycle}
$\Sigma_4$ inside $M_6$.  In order to find a classical D7-brane configuration we have
to solve the equations of motion for both the scalars that parametrize the embedding and
the gauge field living on the brane.  The Poincare symmetry reduces the ten dimensional problem to
six dimensions but the problem is still to complicated to be addressed in its generality.
To find an appropriate embedding in $M_6$ is especially difficult if
the background has a non-trivial NS-NS field $dB=H_3$. The situation, however, greatly simplifies if one considers a supersymmetric solution. The $\kappa$-symmetry analysis for the D7-brane reduces the problem to an Euclidean $D3$-brane
extending along the $4$-cycle $\Sigma_4$. The latter was analyzed in \cite{kappasym}
where it was shown that in order to preserve the background supersymmetries
the embedding and the world-volume gauge field have to satisfy the following conditions:

\begin{enumerate}
	\item The cycle $\Sigma_4$ is holomorphic.
	\item The gauge-invariant 2-form field strength
	      \begin{equation}
	      \F \equiv \varphi^\star(B) + 2 \pi l_s^2 F
	      \end{equation}
	       is of type $(1,1)$.
	\item  Field $\F$ is primitive, i.e. it is orthogonal to the pull-back of the K\"ahler form $J$
				\begin{equation}
	      \varphi^*(J)\wedge \F =0\ .
	      \end{equation}
\end{enumerate}	
	
The last two requirements imply  $\F$ being \emph{anti-self dual}
\begin{equation}
\label{ASD}
\F = - \star_4 \F\ ,
\end{equation}
where the positive orientation on the $4$-cycle $\Sigma_4$ is given by the volume form
\bea
\label{wC}
w_C= \frac{1}{2} \varphi^\star(J \wedge J).
\eea

What happens, however, if we are after a \emph{non-supersymmetric} D7-brane? Unlike the $\kappa$-symmetry condition
above the anti-self duality condition (\ref{ASD}) does not use the complex structure inherited from $M_6$ and
 it is natural to ask whether it can be relevant for a non-supersymmetric case. The answer can be formulated in the form of the following observation.

\emph{
The classical equations of motion for the gauge fields on the D7-brane world-volume are solved by any anti-self dual field $\F$ (\ref{ASD}). Furthermore, the anti-self dual gauge field $\F$ does not contribute to the
equations of motion of the scalars that fix the embedding. In other words, the effective action for the scalars is the
pure DBI action with zero gauge field i.e. the volume of the cycle $\Sigma_4$. Hence any embedding which extremizes the volume, together with any anti-self-dual gauge field on it solves the equations of motion for the probe D7-brane.
}

This is equally applicable to an anti-D7-brane if the anti-self-duality condition is substituted by self-duality.

Before we proceed with the proof let us notice that the observation above provides a novel way to construct classical solutions for D7-brane. At first one can find an embedding $\Sigma_4$ which extremizes the  volume
and later find gauge field configuration which is anti-self-dual and satisfies the Bianchi identity.

The observation directly follows from the fact that \emph{for the given induced metric the D7-brane action is bounded from below and reaches its minimum  if $\F$ is anti-self-dual.}

The rest of this section will prove the observation.
The D7-brane action consists of the two  parts
\begin{equation}
S_{D7} = S_{\rm{DBI}} + S_{\rm{WZ}},
\end{equation}
where
\begin{equation}
\label{DBI-WZ}
S_{\rm{DBI}} = \mu_7 \int \d \sigma^8 e^{-\phi} \sqrt{ -\left \vert \varphi^\star(g_{10}) + \F \right \vert}
\qquad
\textrm{and}
\qquad
S_{\rm{WZ}} = \mu_7 \int \sum_{p} C_p \wedge e^\F\ .
\end{equation}
Here $\mu_7 = (2\pi)^7 l_s^8$, and $\varphi^\star(g_{10})$ is the pullback of the ten dimensional metric (\ref{metric10}).

Let us now show that for the case at hand only the $C_4$ dependent Wess-Zumino term contributes to the action.
First, we have shown above that $F_7=0$. Next, the $C_2$ term vanishes as we assume $\F$ has no no space-time legs due to Poincare symmetry. Finally, the term containing the Hodge dual of $C_4$ is excluded for the same reason as $C_2$.

Now we can plug the explicit form of $C_4$ (\ref{F5C4}) into the WZ action. Furthermore, since we are looking for a static classical solution, we can integrate over the space-time coordinates. The resulting action will depend only
on the $4$-cycle $\Sigma_4$ in $M_6$
\begin{equation}
\label{SgM}
S_{D7} = \mu_7 g^{-1}_s \textrm{Vol}_{\mathbb{R}^{1,3}} \cdot \int_ {\Sigma_4} \d \sigma^4
           \left( \sqrt{\vert g + h^{-1/2}\F \vert} +{\rm Pf}\left(h^{-1/2}\F\right) \right)\ .
\end{equation}
Here $g$ is the induced metric on the $4$-cycle without the warp factor i.e. $\varphi^\star(g_{10})= h^{-1/2} \d x_\mu \d x^\mu + h^{1/2}g$.

To find a classical solution we would have to consider the variation of this action with respect to the world-volume gauge fields $A_\mu$ and the scalars
$\phi_i$ that determine the embedding. The former appear only in $\F$ but the latter in both $g$ and $\F$.
Remarkably, the analysis significantly simplifies if instead of writing down the explicit equations of motion we  try to find the \emph{global} minimum with respect to gauge field $\F$
for the given metric $g$.
Such a minimum indeed exists due to an inequality
\begin{equation}
\label{Smin}
\sqrt{\vert \mathbf{1} + \M \vert} +{\rm Pf}\M \geqslant 1 ,
\end{equation}
where $\M$ is some antisymmetric matrix. It is saturated when $\M$ is anti-self-dual.
An easy way to analyze (\ref{SgM}) is to choose the coordinates such that in a given point $g$ is an identity matrix
and therefore it is invariant under $so(4)\cong su(2) \times su(2)$ symmetry. At this point
$\mathcal{M}=h^{-1/2}\F$ is antisymmetric and the action (\ref{SgM}) can be written in the form (\ref{Smin}).
The antisymmetric $\M$ and can be represented as a sum of self-dual and anti-self-dual parts $\M=\M_s+\M_a$, each in a vector (adjoint) representation of one of the two $su(2)$'s. We denote the length of these two vectors as $r_s$ and $r_s$. These lengths are the only invariants of $\M$ and hence the left-hand-side of (\ref{Smin})
is a function of $r_s$ and $r_a$. We further use $SU(2)\times SU(2)$ to align the ``vectors'' $\M_s$ and $\M_a$
along the third axis
\bea
\M=\left(
\begin{array}{cccc}
0 & (r_s+r_a) & 0 & 0 \\
-(r_s+r_a) & 0 & 0 & 0\\
0 & 0 & 0 & (r_s-r_a)\\
0 & 0 & -(r_s-r_a) & 0
\end{array}
\right) .
\eea
A straightforward calculation leads to inequality (\ref{Smin}) written in the form
\bea
\sqrt{\left(1+(r^2_a-r^2_s)\right)^2+4r^2_s} - (r^2_a-r^2_s) \geqslant 1\  .
\eea
Clearly this inequality is saturated for $r_s^2=0$ and arbitrary $r_a^2$ which means that $\M_s=0$
and so $\M$ is anti-self-dual.

Now we see that the anti-self-dual field minimizes the action for any given $g$ and the minimal value of the action
is simply the geometrical volume. Therefore any embedding which extremizes the volume
\bea
\label{actionvolume}
S_{D7}^{\rm eff} = \mu_7 g^{-1}_s \textrm{Vol}_{\mathbb{R}^{1,3}} \cdot \int_ {\Sigma_4} \d \sigma^4 \sqrt{ \vert g \vert} \ ,
\eea
will solve the full set of the equations of motion if accompanied by any anti-self-dual field $\F$.
If such a gauge field also satisfies the Bianchi identity
\bea
\label{H3cond}
\d\F = \varphi^*(H_3),
\eea
the resulting configuration is a classical solution for the probe D7-brane.

Let us note here that the effective action (\ref{actionvolume}) is calibrated by the integral of $w_C$ (\ref{wC})
over $\Sigma_4$ and the calibration condition is satisfied when the embedding is holomorphic \cite{Calibr}.
In this case ASD solution will satisfy the  $\kappa$-symmetry condition  and be supersymmetric.
At the same time not all ASD solutions are supersymmetric. If embedding merely extremizes
(but does not minimize) the volume, or has to satisfy certain boundary conditions incompatible with
holomorphicity the resulting ADS solution is not supersymmetric. Similarly to \cite{nonBPS} we will call such a solution non-BPS. It satisfies the first order constraint (\ref{ASD}) and minimizes the action with respect to the perturbations
of the world-volume gauge fields.

Moreover one can often easily investigate stability of the ASD solution.
If an embedding (i.e. the resulting induced metric  ${g}$) minimizes the volume (\ref{actionvolume})
within a certain class of geometries (for example invariant under certain symmetry group),
then an anti-self-dual solution ``built'' on top of this embedding is stable within this class of geometries
as the action reaches its minimal value within the class.
If there is more than one anti-self-dual gauge-field which satisfies Bianchi identity  for the given metric ${g}$
the solution might have zero modes associated with the unfixed D3-brane charge.

\section{\bf D7-branes in the KS case}
\label{D7KS}

To illustrate how the observation from Section \ref{secGI} can help one find new solutions we turn to the example of the KS background and construct an explicit example of a non-SUSY probe D7-brane.
As was mentioned above the initial step in finding new solutions is to find an embedding $\Sigma_4$ which extremizes the volume.
Certainly any holomorphic $\Sigma_4$ would do the job, but the resulting ASD solution would necessarily  be supersymmetric.
Alternatively one can focus on $\Sigma_4$ invariant under a sufficiently large group of symmetries which fixes the shape of $\Sigma_4$. Thus in the KS case one can focus on $\Sigma_4$ preserving a part of the global $SO(4)$ symmetry of the conifold. Clearly any D7-brane necessarily breaks $SO(4)$. The biggest unbroken subgroup a D7-brane can preserve is $SU(2)$.
There are two distinct ways how one can embed the unbroken $SU(2)$ into
$SO(4)\cong \left(SU(2)_L\times SU(2)_R \right)/ \mathbb{Z}_2$.
The first one leaves the diagonal $SU(2)_D$ unbroken. The second one preserves only the $SU(2)_R$
(or equivalently the $SU(2)_L$) subgroup.

The $SU(2)_D$ invariant embedding leaves the homogeneous coordinate of the conifold $z_4$
invariant and acts on $(z_1,z_2,z_3)$ as a vector. The lowest state in this sector is a supersymmetric embedding  $z_4={\rm const}$ discussed in \cite{Me}. As was mentioned in the Introduction this solution does not spontaneously break flavor chiral symmetry.

To find the solution preserving one of the two $SU(2)$'s we use the coordinates introduced in \cite{PT}
which make it explicit. Following Papadopoulos and Tseytlin we represent the base of the conifold as the
product $S^2\times S^3$ with one of the $SU(2)$ symmetries acting on the $S^3$ and leaving the $S^2$
invariant while the other $SU(2)$ does not preserve the $S^2\times S^3$ splinting and mixes the two.
Let \bea
\label{t1}
e_1=d\theta_1\ ,\ \ \ e_2=\sin\theta_1 d\phi_1\
\eea
be a basis of one-forms on  $S^2$
and $\epsilon_i$ with $i=1,2,3$ be a basis of the Maurer -- Cartan $SU(2)_R$ invariant forms on $S^3$
\bea
\label{MC}
d\epsilon_i =-\frac{1}{2}\epsilon_{ijk}\epsilon_j\wedge \epsilon_k\ .
\eea
Then the deformed conifold metric written in this notations is \cite{PT,Butti}
\begin{eqnarray}
\label{metric}
ds^2_{M_6} &=& \frac{1}{4} \epsilon^{4/3} K(\tau) \cosh \tau  \Big[ e_1^2 + e_2^2+ \epsilon_1^2 + \epsilon_2^2 + \nonumber \\
           &&  \qquad + \frac{2}{\cosh\tau}(e_1\epsilon_1+e_2\epsilon_2)
                                        + \frac{ 2 }{ 3 K(\tau)^3 \cosh\tau} \left( d\tau^2 + \epsilon_3^2 \right)
               \Big]\ , \nonumber  \\
&&   \textrm{with} \quad K(\tau) = \frac{ ( \sinh\tau\cosh\tau-\tau )^{1/3}}{\sinh\tau},
\end{eqnarray}
while the 10-dimensional metric appears in (\ref{metric10}).
Besides the metric we will also need the NS-NS form
\begin{eqnarray}
B &=&  h_1(\tau)  \left[\epsilon_1\wedge \epsilon_2+e_1\wedge e_2
          +\frac{1}{\cosh \tau}\left(e_1\wedge \epsilon_2 +\epsilon_1\wedge e_2\right) \right]\ , \nonumber\\
&& \textrm{with}   \quad
h_1(\tau) = - \frac{1}{2} \left(g_s M l_s^2\right)  \cdot \frac{\cosh\tau(\tau\cosh \tau-\sinh\tau)}{2 \sinh^2\tau}\ .
\end{eqnarray}
The embedding of the D7-brane  $\Sigma_4$ is stretching along the radius $\tau$ and three other direction on the base of the conifold. To be  $SU(2)_R$ invariant $\Sigma_4$  must completely cover the three-sphere  (which is maped into itself under the action of $SU(2)_R$) and be located at the distinct point(s) at $S^2$.

It is clear from (\ref{metric}) that the deformed conifold is not a direct product $R^+\times S^2\times S^3$.
In principle it is possible to find a trivialization, \emph{i.e.} an appropriate coordinate basis on $S^3$ independent
of the position on $S^2$  \cite{S3S2}.
This approach was adopted in \cite{CobiMe} where a non-SUSY $SU(2)_R$ invariant D7-brane
was found in the background of the singular KW solution. Here we generalize this configuration to the deformed conifold case but there is no apparent need to use the trivialization basis. In fact we are looking for the embedding covering
$S^3$ completely and therefore the details of the coordinate basis on $S^3$ are not important.
This would not be the case if we were after the non-$SU(2)_R$ invariant physics, such as the spectrum of meson perturbations for the model in question.
As we will see below the induced metric on $\Sigma_4$ written in the ordinary coordinates splits into a direct product
$R^+\times S^3$ with the metric on $S^3$ being explicitly $SU(2)_R$ invariant.

As was outlined above we consider an embedding which covers $S^3$ and can be represented as a distinct point
on $S^2$. The location of the latter may depend on the radius $\tau$. Hence the embedding can be visualized as trajectory
$(\phi_1(\tau),\theta_1(\tau))$ on $S^2$. The boundary condition at any given $\tau$ defines a plane which splits $S^2$ into  two equal semi-spheres and preserves the reflection-symmetry through this plane. This implies that the trajectory
$(\phi_1(\tau),\theta_1(\tau))$ is a part of a big circle and with help of the broken $SU(2)_L$ can be aliened along the equator $(\phi_1=\phi(\theta),\theta_1=\pi/2)$\footnote{It can be easily verified that choosing the longitude $(\phi_1={\rm const},\theta_1(\tau))$ would lead to exactly the same induced metric.}.
One can easily see that the resulting unwarped induced metric on $\Sigma_4$ is explicitly $SU(2)_R$ invariant (the dot stands for the $d/d\tau$ derivative)
\begin{eqnarray}
\label{metric4}
ds_{\Sigma_4}^2 &=& \frac{\epsilon^{4/3}}{4} K(\tau) \cosh \tau
\Big[
 \epsilon_1^2 + \tilde{\epsilon}_2^2 +                                    \nonumber \\
&& \qquad  + \left(\tanh^2 \tau \dot\phi^2+ \frac{2}{3K(\tau)^3 \cosh\tau} \right)d\tau^2
 + \frac{2}{3K(\tau)^3 \cosh\tau } \epsilon_3^2
\Big]\ ,                                                                 \nonumber \\
&& \quad \textrm{where} \quad \tilde{\epsilon}_2 \equiv \epsilon_2 + \frac{\dot\phi}{\cosh\tau} d\tau\ .
\end{eqnarray}
After integrating the directions along $S^3$ the effective action (\ref{actionvolume}) becomes
\begin{equation}
\label{effectiveA}
S_{D7}=\mu_7 \frac{2 \pi^2}{24} \epsilon^{8/3} \int \d \tau \frac{\cosh (\tau)}{K(\tau)}
           \left( 1 +  \frac{3 K(\tau)^3 \sinh^2(\tau)}{2 \cosh(\tau)} \cdot  \dot{\phi}^2   \right)^{1/2}\ ,
\end{equation}
which leads to
\bea
\label{emb}
\dot{\phi}^2=\frac{2\cosh\tau}{3K(\tau)^3\sinh^2\tau }
\left( \frac{K(\tau)\sinh^2\tau\cosh\tau}{K(\tau_0)\sinh^2\tau_0\cosh\tau_0 } -1  \right)^{-1}\ .
\eea
This configuration does not necessarily stretches to the bottom of the conifold $\tau=0$
but rather to the minimal radius $\tau=\tau_0$. There the value of $\dot\phi$ jumps from minus to plus
infinity which means that the D7-brane turns back into the direction of larger radius. Hence the geometry is $U$-shaped,
accompanied by a never-shrinking $S^3$ at each radius.

To calculate the total angle ``traveled'' by the trajectory on $S^2$ we integrate $\dot\phi$ from the lowest radius $\tau_0$ to infinity
\begin{equation}
\label{Deltaphi}
\half \Delta \phi = \left( \frac{2}{3} \right)^{1/2} \int_{\tau_0}^\infty \d \tau
                    \left( \frac{\cosh \tau}{K(\tau)^3 \sinh^2(\tau)} \right)^{1/2}
                    \left( \frac{K(\tau) \sinh^2(\tau) \cosh \tau}{K(\tau_0) \sinh^2(\tau_0) \cosh \tau_0} - 1 \right)^{-1/2} \ .
\end{equation}
For a very large $\tau$ we have $K(\tau) \approx 2^{1/3} e^{-\tau/3}$ and one can check that for $\tau_0$ large enough
$\half \Delta \phi = \frac{\sqrt{6}}{8} \pi$ which coincides with the KW case \cite{CobiMe} . On the other hand
for the small values of $\tau$ we have
$K(\tau) \rightarrow (2/3)^{1/3}$ and so in the limit $\tau_0 \to 0$ the integral reduces to
\begin{equation}
\half \Delta \phi = \lim_{\tau_0 \to 0} \int_{\tau_0}^\infty
          \frac{\d \tau }{\tau \left( \left( \frac{\tau}{\tau_0} \right)^2 -1 \right)^{1/2}} = \frac{\pi}{2}.
\end{equation}
This means that for $\tau_0 = 0$ we have an antipodal configuration. Figure 1 
demonstrates the dependence of the
angle $\Delta \phi$ on the parameter $\tau_0$.

\begin{figure}[!h]
\label{Dphi}
\begin{center}
\includegraphics[scale=1.2
]{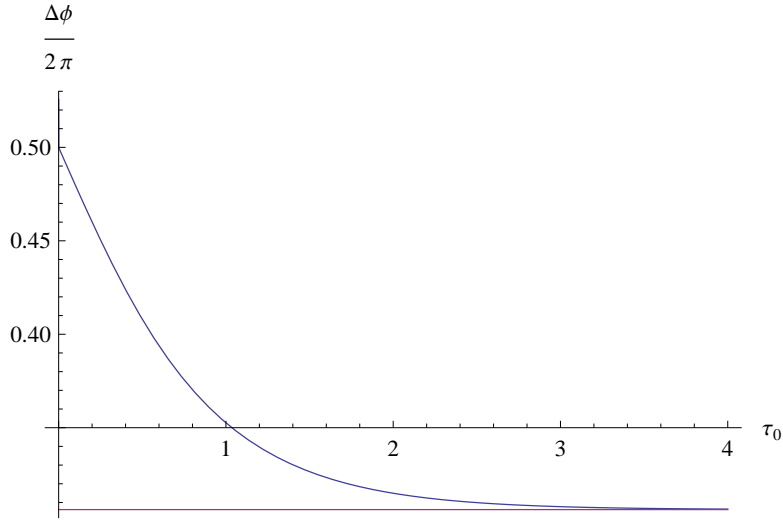}
\caption{The blue plot shows the dependence of the angle $\Delta \phi$ in (\ref{Deltaphi}) on the lowest point of the  profile $\tau_0$. For $\tau_0=0$ the brane and the anti-brane are antipodal ($\Delta \phi=\pi$), whereas
for $\tau_0 \to \infty$ we recover the KW result  $\Delta \phi \rightarrow \sqrt{6}/4 \pi$ (red line).}
\label{Profile2}
\end{center}
\end{figure}

\section{Anti-self-duality condition}
\label{ASDconditionKS}

Our nest task is to complete the embedding (\ref{emb}) by an anti self dual gauge field which satisfies Bianchi identity.
Such a gauge field will minimize the action. Therefore it is reasonable to assume that it preserves the unbroken $SU(2)_R$ symmetry. Hence we proceed with an explicitly $SU(2)_R$ invariant ansatz for the gauge field on $\Sigma_4$ in the  $A_\tau=0$
gauge
\bea
\label{gf}
A(\tau) = A_i(\tau)\epsilon_i\ .
\eea
Given that the pull-back of the NS-NS form is
\bea
\varphi^*(B)=h_1(\tau)\epsilon_1\wedge \tilde\epsilon_2\ ,
\eea and using (\ref{MC}) one can easily calculate the gauge-invariant combination (from here on we  put $g_s Ml_2^2=2$)
\begin{eqnarray}
\nonumber
&& \F = \left( \dot A_3(\tau) + \frac{A_1(\tau)}{\cosh\tau} \dot\phi(\tau)  \right) d\tau \wedge \epsilon_3
             + \left( \dot A_1(\tau) - \frac{A_3(\tau)}{\cosh\tau }\dot\phi(\tau)  \right) d\tau \wedge \epsilon_1  +
 \\
   && \quad  + \dot A_2(\tau) d\tau \wedge \epsilon_2 - (A_3(\tau) - h_1(\tau)) \epsilon_1 \wedge \tilde{\epsilon}_2
       - A_1(\tau) \tilde{\epsilon}_2\wedge \epsilon_3 + A_2(\tau)\epsilon_1  \wedge \epsilon_3.
\end{eqnarray}
Because of the U-like shape of the D7-brane's profile we can not use $\tau$ as a world-volume coordinate globally.
Nevertheless we can start at one branch in the large $\tau$ region and write down
down the anti-self-duality condition (\ref{ASD}) using the induced metric (\ref{metric4})
\begin{eqnarray}
\label{A1A2A3}
\left(\dot A_1(\tau)-\frac{\dot\phi(\tau)}{\cosh\tau} A_3(\tau) \right) &=&  L(\tau) A_1(\tau) , \nonumber\\
\dot A_2(\tau) &=& L(\tau)  A_2(\tau) ,                                                           \nonumber\\
\left(\dot A_3(\tau) + \frac{\dot\phi(\tau)}{\cosh\tau} A_1(\tau) \right) &=&
                                   \frac{2 L(\tau)}{ 3K(\tau)^3\cosh\tau}  (A_3(\tau) - h_1(\tau)) ,
\end{eqnarray}
where
\bea
L(\tau) = \left( 1 + \frac{3K(\tau)^3 \sinh^2\tau}{2  \cosh\tau } \dot\phi(\tau)^2  \right)^{1/2}=
\left( 1 - \frac{ K(\tau_0)\sinh^2\tau_0\cosh\tau_0 }{ K(\tau)\sinh^2\tau\cosh\tau }  \right)^{-1/2}.
\eea
These equations are valid everywhere on that branch from $\tau=\infty$ to $\tau=\tau_0$. When $\tau$ reaches $\tau_0$
and we move to another branch the orientation flips (because $d\tau$ changes sign) and so $L\rightarrow -L$ in the equations above.  To better understand this flip of sign we can switch from $\tau$ to the well-defined coordinate $\phi$ (\ref{emb}). Assuming now
the gauge field  (\ref{gf}) is $\phi$-dependent $A=A(\phi)_i\epsilon_i$ we can write down the ASD equation valid everywhere on the profile. Let us put $A_1=A_3=0$ and focus on $A_2$ for simplicity. The ASD equation in this case takes the form
\bea
\label{asd2}
&&   \qquad \qquad
 \frac{\d A_2(\phi)}{ \d\phi} = \tilde{L}(\phi) A_2(\phi)  \\
&& {\rm with} \quad
\tilde{L}=\frac{L}{ |\dot{\phi}|}=\left(\frac{3K(\tau)^3\sinh^2\tau}{2\cosh\tau} \cdot
\frac{K(\tau)\sinh^2\tau\cosh\tau}{K(\tau_0)\sinh^2\tau_0\cosh\tau_0}\right)^{1/2}
\quad  {\rm for} \quad \tau=\tau(\phi).
\nonumber
\eea
Since $L(\phi)$ and $\tilde{L}(\phi)$ are positively defined and the equation (\ref{asd2}) is valid everywhere on D7-brane we have to conclude that the right-hand-side of the equations (\ref{A1A2A3}) changes sign when $\tau$ moves from one branch into another.
The branch with the plus sign as in (\ref{A1A2A3}) describes the D7-brane while the other branch with minus sign in front of $L$ is the anti-D7-brane.

Let us now analyze the ASD equations. We start with the D7-brane branch i.e. the equations (\ref{A1A2A3}). At large $\tau$ we have $L(\tau) \rightarrow 1$. The function $A_2(\tau)$ is decoupled from the other two functions and therefore it grows like $e^\tau$ if not zero.
This means that we have to put $A_2(\tau)=0$ to satisfy the boundary conditions at infinity.

We were not able to find an analytical solution for $A_1$ and $A_3$ for general $\tau_0$. But for $\tau_0=0$ the derivative $\dot\phi(\tau)$
vanishes everywhere except the origin and the solution can be easily constructed. First we notice that in this case
$A_1(\tau)$ satisfies the same equation as $A_2(\tau)$ in (\ref{A1A2A3}). Therefore the only non-divergent solution is $A_1(\tau)=0$.
A general solution for the remaining equation for $A_3(\tau)$ is
\bea
\label{A3sol}
A_3(\tau) = e^{S(\tau)} \left ( \int_\tau^\infty d \tau^\pr
                            \frac{2h_1(|\tau^\pr|)}{3K(\tau^\pr)^3\cosh \tau^\pr} e^{-S(\tau^\pr)} + c_0 \right)\  \nonumber \\
{\rm with} \quad  S(\tau) =\int_0^\tau d\tau' \frac{2}{3K(\tau')^3\cosh\tau'}  \qquad \qquad
\eea
and $c_0$ is an integration constant.
It is easy to see that $S(\tau) \approx \frac{2}{3} \tau$ for large $\tau$ and so $A_3(\tau)$ exponentially diverges if $c_0$ is not zero.
On the other hand, for $c_0=0$ we find for large $\tau$
\bea
\label{A3asympt}
A_3(\tau) \rightarrow     \frac{-\tau}{2} - \frac{1}{4} .
\eea 
One can also check that $A_3(\tau)$ is regular at $\tau=0$ for any value of $c_0$.

We analyze the equations for $A_1$ and $A_3$ for $\tau_0 >0 $ in Appendix \textbf{B}
and show that there is a unique solution regular at the UV and that it is regular at $\tau=\tau_0$.

To find the solution on the anti D7-brane branch we notice that the homogeneous solution to
(\ref{A1A2A3}) with $-L$ instead of $L$ is a decaying exponent at infinity. Therefore any solution is regular at UV.
The solution is nevertheless uniquely fixed by the boundary condition at $\tau=\tau_0$ to ensure that the solutions on both branches are smoothly glued at the junction. To prove that such a solution exists it is enough to switch to the coordinate $\phi$ like in (\ref{asd2}) since the point $\tau=\tau_0$ is not special in this case.

To illustrate the point that the solution on both branches can be smoothly glued at $\tau_0=0$
we return to the function $A_3(\tau)$ in the special case $\tau_0=0$. In this case we can introduce negative $\tau$ to describe the anti D7 branch while the positive $\tau$ describes the D7 branch\footnote{
\label{zcoord}
For $\tau_0=0$ we can define a new set of coordinates $y=\tau \cos(\phi)$ and  $z=\tau \sin(\phi)$ which are similar to the coordinates introduced in \cite{CobiMe} for the singular conifold. Now the $\tau_0=0$ embedding is simply given by $y=0$ and $z$ is the coordinate that spans the brane worldvolume. Moreover, $z$ ranges from $-\infty$ to $\infty$ and the reflection $z \rightarrow -z$ exchanges the two branches of the brane profile like in \cite{CobiMe}. The radial coordinate is given by $\tau=\vert z \vert$, where the negative values of $\tau$ now correspond to negative $z$. We therefore can rewrite the equation for $A_3$ in  (\ref{A1A2A3}) entirely in terms of $z$ while providing legitimacy to (\ref{A3sol}) for $\tau<0$.}.
This choice is convenient because
(\ref{A3sol}) continued to the negative values of $\tau$ is a solution on the ${\bar D}$7 branch\footnote{Notice the modulus  in the argument of $h_1$ in (\ref{A3sol})}. The solution is obviously smooth at $\tau=\tau_0=0$ and rapidly approaches its asymptotic at plus (\ref{A3asympt}) and minus infinity
\bea
\label{A3asympt2}
A_3(\tau)\rightarrow \frac{-|\tau|}{2}+\frac{5}{4}\ ,
\eea
as shown on Figure 2.

To summarize, the unique choice of the non-divergent boundary conditions for the gauge fields $A_i(\tau)$ in the UV
on the $D$7 branch unambiguously fixes the solution which is regular and smooth everywhere, including IR region $\tau=\tau_0$
and the UV region on the $\bar D$7 branch.

\begin{figure}[!h]
\label{A3plot}
\begin{center}
\includegraphics[scale=1.2
]{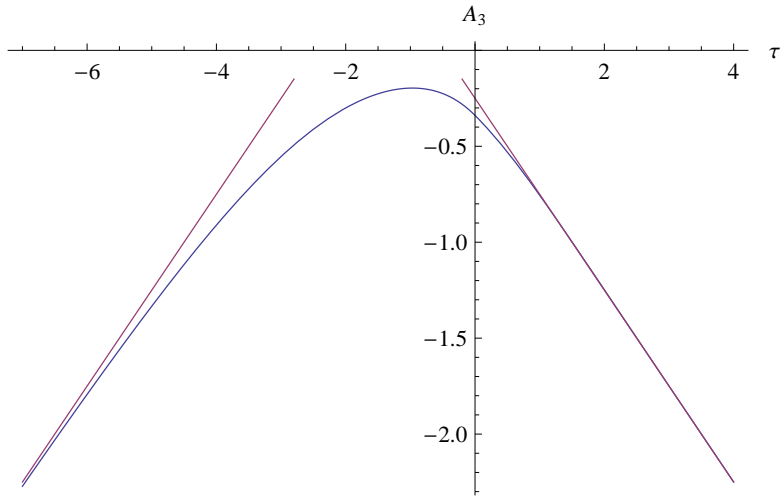}
\caption{The blue plot is the numerical solution for $A_3(\tau)$ for $\tau_0=0$.
The solution rapidly approaches its asymptotic behavior (\ref{A3asympt}) and (\ref{A3asympt2}) (red line).}
\label{Profile1}
\end{center}
\end{figure}

\section{The dual gauge field theory}
\label{dGT}

In \cite{CobiMe} a solution for  $N_{\rm f}$ probe D7 and  anti D7 branes merging in the IR was found in the context of the conformal KW background. In the previous two sections we generalized this result to the KS case.
In this section we will address the phenomenological differences between the two models in question.

\begin{enumerate}
\item{\emph{Broken conformal symmetry}}

				Because of the conformal symmetry of the KW background the asymptotic angle separation
				between the branes was ``frozen" to $\Delta \phi = \frac{\sqrt{6}}{4} \pi$
				and did not depend on the position of the lowest point along the profile $r_0$.
				Therefore the Goldstone boson $\frac{\partial}{\partial r_0}$ associated with
				the spontaneous symmetry breaking of scale invariance did not
                change boundary conditions i.e. was normalizable.
				As a result the massless  Goldstone boson was present in the physical spectrum of the problem.
				This obviously is not the case in the KS model. The KS is a confining background and is obviously
				not scale invariant.  As a result the angle separation $\Delta \phi$ varies
                between $\frac{\sqrt{6}}{4} \pi$ and
				$\pi$ as one takes $\tau_0$ from infinity to zero.
                Therefore the massless mode $\frac{\partial}{\partial \tau_0}$
				changes the boundary conditions, namely not normalizable, and hence not in the spectrum.

\item{\emph{Broken chiral symmetry}}

				As for chiral symmetry we expect that the scenarios in the KW and the KS are quite  similar.
				In particular the underlying gauge theories both admit chiral symmetry which is spontaneously broken by a VEV.
				It was argued in \cite{CobiMe} that the D7 branes described there give rise to the left and right \emph{Weyl} spinors
				rather than the Dirac spinors. The argument was based on the observation that, if taken separately,
				each branch of the brane's profile is similar to one ``half" of the holomorphic embedding of \cite{Ouyang}.
				The latter setup includes Weyl spinors as can be shown using the ``original" $\mathcal{N}=2$ gauge theory \cite{KW}. Since in our case the branches are separated in the UV the Weyl spinors can not interact and hence massless.
				In the IR the chiral symmetry must be broken to the diagonal flavor symmetry due to a quark anti-quark
				condensate.  The spontaneous breaking of the chiral symmetry gives rise to a Goldstone boson which can be
        identified as a massless mode (pion) in the mesonic spectrum on the gravity side.
				From the gauge theory perspective, one can argue that the chiral symmetry must
				be spontaneously broken based on the 't Hooft anomaly matching condition.
				This is similar to the situation in ordinary QCD and for $N_{\rm  f}<N_c$ for ${\cal N}=1$ SQCD.
				
				Let us show that the massless Goldstone mode is present in the spectrum also for the KS background.
				It was argued in \cite{CobiMe} that the $5d$ Maxwell action takes
                the form (here the integral is symbolic and covers both branches)
				\beq \label{Maxwell5d}
				S = -T^\prime \int \d x^4 \d \tau \left( C(\tau) F_{\mu \nu}  F^{\mu \nu} + 2 D(\tau) F_{\mu \tau} F^\mu_\tau \right)
				\eeq
				and a zero mode in the vector spectrum appears if and only if the integral
				\beq
         \label{normalization:eq}
        \int_{\tau_0}^\infty \frac{\d \tau}{D(\tau)}
        \eeq
				is finite. We find the explicit form of the functions $C(\tau)$ and $D(\tau)$ in Appendix \textbf{C}.
				It appears that $D(\tau)$ does not depend on the gauge fields on the $S^3$
                and therefore is the same on both branches
				\beq
				D(\tau)=\vert g \vert^{1/2} g^{\tau\tau}
                       = \frac{\sqrt{3}}{4} \epsilon^{4/3} \frac{K(\tau) \cosh(\tau)}{L(\tau)}\ .
				\eeq
				 This expression is finite for $\tau \to \tau_0$ since $D(\tau_0)$ behaves like $(\tau-\tau_0)^{1/2}$ near $\tau_0=0$. Therefore the integral (\ref{normalization:eq}) is obviously convergent.

\item{\emph {Broken supersymmetry}}

        It was shown in \cite{CobiMe} that the two branches of the D7-brane profile asymptotically can be described via
        \bea
        \label{2br}
        {w_1\over w_3}=\lambda_1\ ,\ \ \ \ {w_1\over w_3}=\lambda_2\ ,\ \ \ \
        {\lambda_1\over \lambda_2}=e^{i\Delta\phi}\ .
        \eea
        Here $w_i$ are the homogeneous coordinates on the singular conifold
        \bea
        w_1 w_2-w_3 w_4=0\ ,
        \eea
        and the two complex constants $\lambda_1,\lambda_2$ mark the location of the branches on $S^2$.
        Although each branch is asymptotically holomorphic together they do not lie on a holomorphic curve.
        Therefore the corresponding solution is not supersymmetric.  Since the geometry of the deformed confiold approaches the
        singular conifold in the UV the description (\ref{2br}) is still valid in the KS case. Therefore the solution                   found in the previous sections is not supersymmetric.  From the field theory perspective the broken supersymmetry implies that while the chiral quarks are massless, the corresponding flavor scalars (these would be squarks in a supersymmetric case) are massive.  It also implies that the fermionic spectrum on the flavor brane should not admit a massless Goldstino. We expect this can be proved by a careful analysis of the	D3-D7 strings.
\end{enumerate}

Another important question that we would like to address is CP violation.
We saw in the previous section that the ASD world-volume flux does not respect the $\mathbf{Z}_2$ symmetry that
exchanges the two branches of the D7-brane (i.e. $\phi \to -\phi$). We argue that this $\mathbf{Z}_2$ breaking
implies the C and P-symmetry breaking while CP is unbroken.

Before we turn to the analysis of the KS case let us review the results derived in \cite{CobiMe} for D7-brane in the KW background.
In this case the world-volume flux vanishes and the embedding is symmetric under $\mathbf{Z}_2$. It was shown there that both C and P transformations involve the $\mathbf{Z}_2$ flipping $\phi \to -\phi$. Yet one
can assign charges to all fields such that the effective five and four-dimensional actions are C and P invariant.
To illustrate the point we consider a five-dimensional  interaction term
\bea
\label{intterm}
\int d^4 x\ d\phi \ \delta\theta_1\ F\wedge F\ ,
\eea
which follows  from the WZ term $\int_{D7}\tilde{C}_4e^{\F}$. Here $\delta\theta_1$ is the KK mode of $\theta_1$
defined in (\ref{t1}). Since C (and P) turn $\phi$ into $-\phi$ for all fields in (\ref{intterm})
the expression is invariant no matter how the wave-functions $F$ and $\delta\theta_1$ look like.
The transformation $\phi \to -\phi$ is simply a change of integration variable from the five-dimensional point of view.
Furthermore since the embedding is $\mathbf{Z}_2$ invariant each individual world-volume wave-function has definite $\mathbf{Z}_2$ charge
and therefore one can assign each fluctuation with a definite C (and P) charge.
This is not so if there is a non $\mathbf{Z}_2$-invariant world-volume flux present, like in the case when D7 is embedded into KS background. It is still true that the interaction term (\ref{intterm}) is invariant under C and P symmetry. But because the individual fluctuations satisfy a non-$\mathbf{Z}_2$ symmetric equation they don't carry a definite charge under $\mathbf{Z}_2$ and hence under C or P. Therefore both C and P symmetry is broken in four dimensions. Despite that the CP symmetry is not violated as the combination of C and P does not act on the fifth coordinate $\phi$.

We leave it as an open problem to find an explicit equation for $\delta \theta_1$ (and other fields)
and estimate the amount of C and P violation. In any case our scenario significantly differs
from the Peccei-Quinn model where CP is broken spontaneously by a non-zero VEV.

Let us finish this section with a  comment on possible applications of our result to the physics of baryons
inspired by the Sakai-Sugimoto model. It was shown in \cite{Sakai:2004cn} that the D8-brane warping the $S^4$-cycle
gives rise to a five-dimensional gauge theory. The later has  solutions with a non-zero baryon charge that correspond to the dissolved D4-branes warping $S^4$. On a more technical level the corresponding solutions are the five-dimensional instantones
which do not shrink to the zero size due to a non-zero Chern-Simons term.
Naturally our model has a potential for a similar construction. Indeed, in our case the D7-brane is wrapping the $S^3$-cycle
and leads to a five-dimensional gauge theory if these directions are integrated out. We expect that a dissolved
D3-brane that is wrapping $S^3$ will similarly give rise to a baryon configuration.
We leave the detailed form of the five-dimensional action and other aspects of the baryon solution for future investigation.

\section{Discussion}
\label{Discuss_s}

In this paper we proposed a new method to construct a non-BPS solutions for the probe D7-brane in the context of the type IIB backgrounds with a constant dilaton. Our method is based on the anti-self-duality condition for the world-volume gauge field which generalizes the $\kappa$-symmetry condition for the SUSY D-brane. We apply this method to the Klebanov-Strassler background and found a non-SUSY solution with a $SU(2)$ symmetry. The resulting profile is a product of the Minkowsi space and  a  U-like shape accompanied at each point by a never-shrinking $S^3$. It can be thought of as a junction of a D7 and an anti D7-brane. Because of the merging region in the IR the asymptotic flavor symmetry $SU(N)_L\times SU(N)_R$ is spontaneously broken down to $SU(N)_D$ by a non-zero VEV. We identified the corresponding massless Goldstone boson in the physical spectrum. We expect our model to provide a holographic realization of the QCD-like physics (mesons and baryons) in a similar way to the Sakai-Sugimoto model. At the same time our model does not suffer from the divergent dilaton in the UV .

Besides describing flavor physics in the QCD-like theories the D7-brane plays a crucial role of stabilizing K\"ahler moduli in the throat compactification models proposed in \cite{KKLT}.
The stringy models of inflation based on this setup include a mobile D3-brane \cite{KKLMMT} with its dynamics
governed by the D3-D7 interaction \cite{d3d7}. It was recently shown that such a model is capable of describing
cosmological inflation matching current experimental date \cite{explicit}. Given the multitude of possible locations
of the D7-brane within the throat the dynamics of the model can significantly vary. The solution for the D7-brane found
in this paper is rather peculiar in this sense as it preserves a large symmetry group and this significantly simplifies
the analysis. Because of the $SU(2)_R$ invariance the only
possible operators perturbing the effective potential of the inflaton are those with quantum numbers $j_2=R=0$
and arbitrary $j_1$. The analysis in \cite{holographic} reveals that in this case instead of the generic form
\bea
V_{D3-D7}(r)\cong c_1 r+c_{3/2}r^{3/2}+c_2 r^2+O(r^4)+..\ , \ \ \
c_i\sim 1\ ,
\eea
the D3-D7 potential acquires a specific form \bea
V_{D3-D7}(r)\cong c'_2 r^2 +O(r^4)+..\ ,
\eea
with a possibility to fine tune the mass parameter $c_2'$ to zero.
It would be interesting to study this model in more detail to investigate if it yields more favorable phenomenological predictions.

In conclusion let us note that the method of constructing non-supersymmetric non-BPS solutions discussed in Section \ref{secGI} can be applied to the interesting problem of holographic description of the SUSY breaking states in supersymmetric field theories. To
construct such a state one has to find an embedding for the D7-brane which satisfies some ``holomorphic'' boundary condition in UV and solves the EOM for the effective action (\ref{actionvolume}) but is not holomorphic. If this solution happens to be metastable it can be used in various applications including the holographic gauge mediation models like the one discussed in \cite{HGM}.

\section*{Acknowledgments}
We thank O.~Aharony, D.~Persson, S.~Kachru, R.~Kallosh, I.~Klebanov, L.~McAllister and S.~Cremonesi for useful discussions.

A.D. acknowledge the hospitality of Weizmann Institute while this work was initiated.
The research of A.D. is supported by the Stanford Institute for Theoretical Physics and
also in part by grant RFBR 07-02-00878, and Grant for Support of Scientific Schools NSh-3035.2008.2.

This research of S.K. is supported in part by the Belgian Federal Science Policy Office through the Interuniversity Attraction Pole IAP VI/11, by the European Commission FP6 RTN programme MRTN-CT-2004-005104 and by FWO-Vlaanderen through
project G.0428.06.

The work of J.S. was supported in part by a center of excellence supported
by the Israel Science Foundation (grant number 1468/06), by a grant (DIP H52)
of the German Israel Project Cooperation, by th BSF grant 2006157, by
the GIF grant 94/2007 and by the European
Network MRTN-CT-2004-512194.

\section*{\bf Appendix}

\addcontentsline{toc}{section}{Appendix}

\subsection*{{\bf A} \ The minimum of the D7-brane action in the case of a general metric}

In this appendix we analyze the action (\ref{SgM}) ans show that it is bounded from below by (\ref{actionvolume}). The inequality is saturated if $\F$ is ASD.

Let $\M$ be a $4 \times 4$ matrix defined by\footnote{Here we ignore the warp function $h$
which can be trivially absorbed in $g$ or $\F$.}
$\M=g^{-1}\F$, where $g$ and $\F$ are arbitrary real symmetric and
anti-symmetric $4 \times 4$ matrices respectively and additionally $g$ is \emph{positive definite}.
Using this notations we can rewrite the essential part of (\ref{SgM}) as
\bea
\sqrt{|g|}\left[\sqrt{|1+\M|}-\sqrt{|\M|}\right]\ .
\eea
From the symmetry properties we immediately find that $\M^\dagger = - g \M g^{-1}$ and so
the matrices $\M^\dagger$ and $(-\M)$ are similar. As such they have the same eigenvalues and the same
(block) diagonal form. Let us denote by $D$ the (block) diagonalized form of
$\M$ and by $P$ a non-singular matrix satisfying $D=P \M P^{-1}$. We then find that
$D^\dagger = - \left( {P^\dagger}^{-1} g P^{-1} \right) D \left( P g^{-1} P^\dagger \right)$.
Since $D^\dagger$ and $(-D)$ have the same diagonal elements up to some re-ordering, the matrix $g$ is necessarily
\emph{congruent} to a permutation matrix. The only permutation matrix, however,
congruent to a \emph{positive definite} matrix is the unit matrix. We conclude therefore that the eigenvalues of
$D$, and $\M$, are purely imaginary. Notice, however, that $\M$ is by definition real and so if $i \lambda$
is an eigenvalue then also is $-i \lambda$. To summarize, the four eigenvalues of $\M$ are $\pm i \lambda_1$
and $\pm i \lambda_2$ for real $\lambda_{1,2}$.
As a consistency check we can verify that $\M$ is indeed traceless as it follows from its definition.
(Noticing further that $(-\M)$ and $\M^T$ are also similar we can easily show that $\M$ is actually diagonalizable and
not block-diagonalizable, but this is not necessary for our discussion).

Using these results we can write:
\begin{equation}
\label{1+M}
\sqrt{\vert \mathbf{1}+\M \vert} - \sqrt{\vert \M \vert} = \sqrt{ (1 + \lambda_1^2)(1 + \lambda_2^2) }
                              - \vert \lambda_1 \lambda_2 \vert \geqslant 1,
\end{equation}
where the inequality is saturated \emph{if only if} $\lambda_1 = \lambda_2$. In this case all the eigenvalues of $\M^2$
are equal and so $\M^2$ is proportional to the unity matrix:
\begin{equation}
\label{Mmin}
\M^2 = - \vert \M \vert^{1/2} \cdot \mathbf{1}.
\end{equation}
In terms of the original matrices $g$ and $\F$ it implies:
\begin{equation}
\label{eFstarF}
\eps^{abcd} \left( \star_4 \F \right)_{ab} \F_{ce} = - 2 \textrm{Pf} (\F) \cdot \delta^d_e,
\end{equation}
where $\star_4$ stands for the Hodge duality with respect to the $4$-dimensional metric $g$.
Comparing this with the identity:
\begin{equation}
\label{eFF}
\eps^{abcd} \F_{ab} \F_{ce} =  2 \textrm{Pf} (\F) \cdot \delta^d_e,
\end{equation}
which holds for any $4 \times 4$ anti-symmetric matrix $\F$, we immediately\footnote{
Going from (\ref{eFstarF}) to (\ref{eFF}) we assumed that $\vert \F \vert \neq 0$. Otherwise
for $\vert \F \vert = 0$
we have $\vert \M \vert = 0$  an so from (\ref{Mmin}) one gets $\M^2=0$. This in turn means
that the matrix $\F g^{-1} \F$ is zero, which is possible only for $\F=0$ and thus
the anti-self duality condition still holds.} arrive at the anti self-duality condition (\ref{ASD}).

\subsection*{{\bf B} \ The solution of the ASD condition (\ref{A1A2A3}) for $\tau_0 > 0$}

Here we consider the solution of the  system of equations (\ref{A1A2A3}) for $\tau_0 > 0$.
In this case $\dot \phi \neq 0$ (or, more precisely, $\dot \phi (\tau)$ is not a $\delta$-function)
and we cannot put $A_1(\tau)=0$ anymore. On the
other hand we still have to put $A_2(\tau)=0$ to avoid the divergence at infinity. If we denote the vector $(A_1(\tau),A_3(\tau))^{T}$ by $A(\tau)$ then the equations
(\ref{A1A2A3}) can be written as:
\bea
\dot{A}(\tau) = Q(\tau) A(\tau) + R(\tau),
\eea
where the $2 \times 2$ matrix $Q(\tau)$ and the $2$-vector $R(\tau)$ can be easily read from (\ref{A1A2A3}). The solution
of this differential equation is:
\bea
A (\tau) = e^{S(\tau)} \left(  \textrm{P} \int_\tau^{\infty} e^{-S(\tau^\pr)} R(\tau^\pr) \d \tau^\pr + C_0 \right) .
\eea
Here $S(\tau)$ is defined by $S^\pr (\tau) = Q(\tau)$ and the integration constant can be absorbed in a proper redefinition
of the constant vector $C_0$. The ordering in the solution appears because of the matrix nature of the equation.  For $\tau \to \infty$ we have:
\bea
Q(\tau) \approx \left( \begin{array}{cc} 1 & 0 \\ 0 & \frac{2}{3} \end{array}\right)
\qquad \textrm{and} \qquad
R(\tau) \approx \left( \begin{array}{c} 0 \\ \frac{\tau}{3} \end{array}\right).
\eea
Thus if $C_0 \neq \left( 0,0 \right)^T$ we have $A_1(\tau) \approx e^\tau$ or $A_3(\tau) \approx e^{\frac{2}{3}\tau}$. Therefore we have to put $C_0=\left( 0,0 \right)^T$ so that at infinity $A_1(\tau)$ vanishes and
$A_3(\tau) \approx - \frac{\tau}{2}$. Now we have to verify that
the solution in question is regular at $\tau=\tau_0$. We find near $\tau_0$
\bea
Q(\tau) \approx \frac{1}{(\tau-\tau_0)^{1/2}}  Q_0
\qquad \textrm{and} \qquad
R(\tau) \approx \left( \begin{array}{c} 0 \\ \frac{1}{(\tau-\tau_0)^{1/2}} \end{array}\right) R_0\ .
\eea
for some non-zero $R_0$ and non-singular $Q_0$. This implies that for small $\tau-\tau_0$, \mbox{$S(\tau) \approx (\tau-\tau_0)^{1/2} S_0$}. Finally both $A_1(\tau)$ and $A_3(\tau)$ behave like ${\rm const}+(\tau-\tau_0)^{1/2}$ at $\tau = \tau_0$ and thus the solution is regular in both  UV and the IR. Clearly this solution can be continued to the second branch. Since the solution to the homogeneous system  behaves like $A_1(\tau) \approx e^{-\tau}$ or $A_3(\tau) \approx e^{-\frac{2}{3}\tau}$ at infinity the solution in question is regular everywhere.

\subsection*{{\bf C} \ The general form of the $5d$ Maxwell action}

In this appendix we will derive the general form of the $5d$ effective action of the gauge fields.

To this end we have to expand the D7 brane action (\ref{DBI-WZ}) around the classical profile.
The perturbation we are interested in is $\delta \F = 2 \pi l_s^2 F$ where $F$ has legs only along the $4d$ space-time and
the radial coordinate $\tau$. As usual the contribution of the WZ part is proportional to $F \wedge F$ and
does not contribute to the equations of motion for an abelian $\F$. After the integration
over the $3$-sphere the variation of the DBI part yields
\bea
\delta S_{DBI} = -T^\pr \int d^4 x d \tau \sqrt{\left\vert E \right\vert} {\rm Tr} \left( (E^{-1} F)^2 \right)
\eea
with $E \equiv \varphi^\star(g_{10}) + \F_0$, where $\varphi^\star(g_{10})$ the pullback of the $10d$ metric $g_{10}$ (\ref{metric10}) and $\F_0$ is the ASD $2$-from we found in Section \ref{D7KS}.
Here we absorbed various numerical and dimensionful constants inside $T^\pr$
and use the fact that ${\rm Tr} (E^{-1} F)$ vanishes. The latter follows from the fact that $\delta \F$ has legs only along the space-time while $\F_0$ has no such legs.
The $8 \times 8$ matrix $E$ consists of two $4 \times 4$ blocks. The first block corresponds to the $4d$ space time
and the second is related to the $4$-cycle of the deformed conifold warped by the D7 brane.
The $8 \times 8$ matrix $E$ can be written schematically as
$E=( h^{-\half} \eta_{\mu \nu}, h^\half g (1+\M_0))$ where the entries correspond to the two blocks. As was already
explained after equation (\ref{SgM}) in our notations $g$ is the un-warped induced metric on the $4$-cycle,
namely $\varphi^\star(g_{10})= h^{-1/2} \d x_\mu \d x^\mu + h^{1/2}g$ and $\M_0 = h^{-1/2} g^{-1} \F_0$.
In the same notation we have $F=(F_{\mu \nu}, F_{\mu \tau})$.

It was demonstrated in Appendix {\textbf A} that the ASD classical solution we are investigating
satisfies (\ref{Mmin}). This in turn implies that
\bea
(\mathbf{1}+\M_0)^{-1} = \frac{(\mathbf{1}-\M_0)}{\left( 1 + \sqrt{\vert \M_0 \vert} \right)}.
\eea
Using this result we find
\bea
E^{-1} = \left( h^\half \eta^{\mu \nu},
             \frac{h^{-\half} \sqrt{\vert g \vert}}{ \sqrt{\vert g \vert} + h^{-1}  \sqrt{\vert \F_0 \vert}}
                                            \left( g^{-1} - h^{-\half} g^{-1} \F_0  g^{-1} \right)  \right).
\eea
and
\bea
\sqrt{\vert E \vert} = \sqrt{\vert g \vert} + h^{-1} \sqrt{\vert \F_0 \vert}.
\eea
Finally, noticing that $\left( g^{-1} \F_0  g^{-1} \right)^{\tau \tau} = 0$, we arrive at the following $5d$
Maxwell action
\bea
S_{5d} = - T^\pr \int d^4 x d \tau \left[
                   \left( h \sqrt{\vert g \vert} + \sqrt{\vert \F_0 \vert} \right) F_{\mu \nu} F^{\mu \nu}
                 + 2  \sqrt{\vert g \vert} g^{\tau \tau} F_{\mu \tau} F^\mu_\tau \right].
\eea


\end{document}